\documentclass[aps,pre,onecolumn,groupedaddress,nofootinbib,preprint,superscriptaddress]{revtex4-1}

\usepackage[utf8]{inputenc}
\usepackage[T1]{fontenc}
\usepackage{cmap}

\usepackage{graphicx}
\usepackage{amsmath}
\usepackage{amsfonts}
\usepackage{amssymb}
\usepackage{xcolor}
\usepackage[unicode,
  pdftitle={Insight into the Li2CO3-K2CO3 eutectic mixture from classical molecular dynamics: thermodynamics, structure and dynamics},
  pdfauthor={Dario Corradini; François-Xavier Coudert; Rodolphe Vuilleumier}]{hyperref}

\usepackage{fancyhdr}
\fancypagestyle{plain}{
\fancyhf{}

\fancyhead[C]{\href{http://dx.doi.org/10.1063/1.4943392}{\textcolor{blue}{Published as: \emph{J. Chem. Phys.} \textbf{144}, 104507 (2016), DOI: 10.1063/1.4943392}}}
}
\begin{document}

\title{Insight into the Li$_{\text 2}$CO$_{\text 3}$--K$_{\text 2}$CO$_{\text 3}$ eutectic mixture from classical molecular dynamics: thermodynamics, structure and dynamics}

\author{Dario Corradini}
\affiliation{Department of Chemistry, \'Ecole Normale Sup\'erieure - PSL Research University, 24 rue Lhomond,
75005 Paris, France}
\affiliation{Sorbonne Universit\'es, UPMC Univ. Paris 06, PASTEUR, 75005 Paris, France}
\affiliation{CNRS, UMR 8640 PASTEUR, 75005 Paris, France}
\author{Fran\c{c}ois-Xavier Coudert}
\email{Email: fx.coudert@chimie-paristech.fr; Twitter: @fxcoudert}
\affiliation{Chimie ParisTech, PSL Research University, CNRS, Institut de Recherche de Chimie Paris, 75005 Paris, France}
\author{Rodolphe Vuilleumier}
\email{Email: rodolphe.vuilleumier@ens.fr; Twitter: @RodVuilleumier}
\affiliation{Department of Chemistry, \'Ecole Normale Sup\'erieure - PSL Research University, 24 rue Lhomond,
75005 Paris, France}
\affiliation{Sorbonne Universit\'es, UPMC Univ. Paris 06, PASTEUR, 75005 Paris, France}
\affiliation{CNRS, UMR 8640 PASTEUR, 75005 Paris, France}

\date{\today}

\makeatletter
\def\frontmatter@preabstractspace{3.0\baselineskip}
\makeatother

\begin{abstract}
\noindent
We use molecular dynamics simulations to study the thermodynamics, structure and dynamics of the Li$_2$CO$_3$--K$_2$CO$_3$ (62:38 mol\%) eutectic mixture. We present a new classical non-polarizable force field for this molten salt mixture, optimized using experimental and first principles molecular dynamics simulations data as reference. This simple force field allows efficient molecular simulations of phenomena at long timescales. We use this optimized force field to describe the behavior of the eutectic mixture in the 900--1100~K temperature range, at pressures between 0 and 5~GPa. After studying the equation of state in these thermodynamic conditions, we present molecular insight into the structure and dynamics of the melt. In particular, we present an analysis of the temperature and pressure dependence of the eutectic mixture's self diffusion coefficients, viscosity and ionic conductivity.
\end{abstract}

\maketitle
\thispagestyle{plain}

\section{Introduction}

Carbonate melts are liquids with remarkable physical and chemical properties that have recently received much attention~\cite{gaillard2008, jones2013}. In addition to being present in Earth's upper mantle, they have also been increasingly used in various technological applications, either as bulk material or in composites with solid oxides or metals. Molten carbonates are successfully used as electrolytes in {\it molten carbonate fuel cells} (MCFC) and are experimented in carbon capture and storage devices. Yet, despite the importance of carbonate melts for both geophysical settings and industrial applications, a clear picture of the structure, dynamics and reactivity of molten carbonates at the molecular scale is still lacking. Here, we set out to use molecular simulation methods to provide insight into the microscopic structure and dynamical properties of the $\rm Li_2CO_3-K_2CO_3$ eutectic mixture (62:38 mol\%), which is of particular relevance for industrial applications as it is a standard electrolyte for molten carbonate fuel cells due to its high ionic conductivity and low eutectic temperature~\cite{li2011}.

{\it First principles molecular dynamics} (FPMD) simulations are one possible choice of simulation technique in order to simulate the carbonate melts. Relying on quantum chemical calculations of the electronic structure at the {\it density functional theory} (DFT) level, they do not require \emph{a priori} knowledge of the molecular interactions in carbonate melts. FPMD simulations have been used with great success in recent years to calculate the equation of state of molten $\rm CaCO_3$~\cite{vuilleumier2014} and the speciation and transport of $\rm CO_2$ in $\rm CaCO_3$~\cite{corradini2015}. However, first principles techniques are computationally very expensive, and therefore inherently strongly constrained on the time scales and system sizes that can be investigated in that manner. On the other hand, molecular dynamics (MD) based on empirical force fields are less computationally demanding, but requiring careful parametrization. For our system of interest, we note that there have been so far relatively few attempts to derive force fields for molten alkali carbonates~\cite{tissen90,habasaki90,costa08}. Therefore, in this work we propose an optimized force field for the $\rm Li_2CO_3-K_2CO_3$ eutectic mixture, based on both experimental data and results from FPMD simulations. We then use this force field to study the thermodynamics, structure and dynamics of the molten mixture in the 900--1100~K temperature range, at pressures up to 5~GPa. We focus in particular on the dynamical properties including the self diffusion coefficients, the viscosity and the ionic conductivity.

\section{Methods}

\subsection{First-principles molecular dynamics}

In order to be able to provide reference structural information to optimize and validate our classical force field, we have first performed FPMD simulations on the $\rm Li_2CO_3-K_2CO_3$ (62:38 mol\%) eutectic mixture at the three target temperatures, $T=900$~K, $T=1000$~K and $T=1100$~K. We conduct these simulations using DFT and the Born--Oppheneimer dynamics. We use the {\footnotesize CP2K} software package~\cite{cp2k} and in particular the {\footnotesize QUICKSTEP} algorithm~\cite{vandevondele2005a}, which employs a hybrid Gaussian plane-wave method (GPW)~\cite{lippert1997} for the electronic structure. The core electrons are replaced by using norm--conserving Goedecker--Teter--Hutter (GTH) pseudo-potentials~\cite{goedecker1996,hartwigsen1998,krack2005}. For all atomic species we use a double-zeta valence plus polarization (DZVP) basis set optimized for molecules~\cite{vandevondele2007}. We cut off the electronic density at 700~Ry and use a \texttt{SPLINE3} smoothing for applying the exchange--correlation potential. The latter is described thanks to the BLYP functional~\cite{becke1988,lee1988}, with added dispersive interactions through the DFT-D3 scheme~\cite{grimme_d3} with a cutoff distance of 40~{\AA}.

The system studied by first-principles simulations is composed of 128 $\rm CO_3^{2-}$, 159 $\rm Li^+$ and 97 $\rm K^+$, with periodic boundary conditions. We adjust the box length in order to prepare the system at the experimental density at atmospheric pressure~\cite{janz88} and we run the simulations in the $NVT$ ensemble. The temperature is controlled by using the CSVR thermostat~\cite{bussi2007} with a time constant of 1~ps. We employ a simulation time step of 0.5~fs and the total simulation time is 30~ps. The initial configurations had been obtained by classical MD using the Tissen and Janssen force field~\cite{tissen90}.

\subsection{Force field}\label{sec:ff}

Relatively few attempts have been previously made to derive a force field for molten alkali carbonates~\cite{tissen90,habasaki90,costa08}. The force field developed by Tissen and Janssen~\cite{tissen90} for pure lithium, sodium and potassium carbonate salts, and later also used to study mixtures~\cite{tissen94,koishi00}, treats the carbonate molecule as rigid. In this simple force field, the interaction potential between the atoms is simply given by the sum of the Coulombic potential and a Born--type repulsive term. Partial charges are used for the atoms composing the $\rm CO_3^{2-}$ units, while formal charges are used for the cations. Costa and Ribeiro~\cite{costa08} have compared the results obtained using the force field by Tissen and Janssen to the ones produced by a force field in which the polarizability of the anion is added in the form of fluctuating partial charges, showing a large impact on the prediction of transport properties.

Here, in light of the length of the runs needed to converge the calculation of dynamical quantities such as diffusion coefficients, ionic conductivity and viscosity, we limit ourselves to a non--polarizable force field. The complexity of the force field used is however somewhat increased with respect the Tissen and Janssen model. For the derivation of our classical force field for $\rm Li_2CO_3-K_2CO_3$, we start from the force field obtained for $\rm CaCO_3$ in Ref.~\cite{vuilleumier2014}. We consider flexible $\rm CO_3^{2-}$ ions. The C--O pairs in carbonate interact via harmonic terms:
\begin{equation}
U_{\rm C-O}^{\rm harm} = \sum_{i \in {\rm C}} \sum_{j \in {\rm O}}^{\rm intra} \frac{1}{2}k_{\rm C-O} (r_{ij}-r_{\rm C-O}^*)^2
\end{equation}
with $k_{\rm C-O} = 6118.17$~kJ/(mol \AA$^2$), $r_{\rm C-O}^*=1.16$~{\AA}, and where ``intra'' means that the sum is taken only for O in the same carbonate molecule of the corresponding C. The carbonate molecule is prevented from folding by adding exponential repulsion terms between O atoms of the kind:
\begin{equation}
U_{\rm O-O}^{\rm rep} = \sum_{i \in {\rm O}} \sum_{j \in {\rm O},\, j>i}^{\rm intra} B_{\rm O-O} \exp (-r_{ij}/\rho_{\rm O-O})
\end{equation}
with $B_{\rm O-O}=2.6117\ 10^6$~kJ/mol and $\rho_{\rm O-O} = 0.22$~{\AA}.

The non--bonded (or intermolecular) part of the force field includes a Coulombic part, a repulsive term and a dispersive term, with its analytical form being:
\begin{equation}\label{eq:nb}
U_{\rm NB}= \sum_{i<j}^{\rm NB} \frac{q_i q_j}{r_{ij}} + B_{ij} e^{-r_{ij}/\rho_{ij}} - \frac{C_{ij}}{r_{ij}^6}
\end{equation}
where ``NB'' means that the sum is taken only over non-bonded pairs of atoms. We use the same partial charges as in Ref.~\cite{vuilleumier2014}, that is to say $q_{\rm C}=+ 1.04085\, e$, $q_{\rm O}=- 0.89429\, e$ and  $q_{\rm Li}= q_{\rm K}=+ 0.82101\, e$. We had initially attempted to reoptimize these charges, but any significant change leads to drastic deterioration of the pressures and densities of the system when compared to experimental data. We thus proceed to adjust the interaction parameters $B_{ij}$, $\rho_{ij}$ and $ C_{ij}$ by manual tuning in order to reproduce at best both the experimental density by constant-pressure $NPT$ simulations and the pair radial distribution functions (RDFs), as obtained from first-principles molecular dynamics (FPMD) simulations, by constant-volume $NVT$ simulations. The best values of the parameters are reported in Table~\ref{tab:1}. All parameters not shown ($B_{\rm Li-Li}$, $C_{\rm Li-Li}$, $B_{\rm K-K}$, $C_{\rm K-K}$, $B_{\rm Li-K}$, $C_{\rm Li-K}$) are equal to zero.

\begin{table}[htbp]
\caption{Non--bonded force field parameters $B_{ij}$, $\rho_{ij}$ for the repulsive and $ C_{ij}$ for the dispersive terms of the interaction potential, Eq.~\ref{eq:nb}. All pair parameters not shown are set to zero.}
\begin{center}
\begin{ruledtabular}
\begin{tabular}{lccc}
Pair & $B_{ij}$ (kJ/mol) & $\rho_{ij}$ (\AA) & $ C_{ij}$ (kJ \AA$^6$/mol)  \\
\hline
Li--O & 15.0 10$^5$ & 0.175 & 0\\
K--O  & 7.0 10$^5$ & 0.249 & 3000.0\\
O--O & 5.0 10$^5$ & 0.253 & 2300.0
\end{tabular}
\end{ruledtabular}
\end{center}
\label{tab:1}
\end{table}

\subsection{Validation of the force field parameters}\label{sec:ffvalid}

We study here the $\rm Li_2CO_3-K_2CO_3$ eutectic mixture (62:38 mol\%) at three temperatures, namely $T=900$~K, $T=1000$~K and $T=1100$~K. The experimental density at atmospheric pressure for the eutectic mixture is given by the formula $\rho= a-b \cdot T$, with $a=2.3526$~g/cm$^3$ and $b=4.532\ 10^{-4}$~g/(cm$^3$ K) in the interval $T=843-1218$~K~\cite{janz88}. This gives at the target temperatures $\rho(900~{\rm K})=1.94$~g/cm$^3$, $\rho(1000~{\rm K})=1.90$~g/cm$^3$ and $\rho(1100~{\rm K})=1.85$~g/cm$^3$. We perform constant-pressure molecular simulation runs of 0.5~ns at atmospheric pressure and at the target temperatures. The temperature is controlled by the Nos\'e-Hoover thermostat~\cite{hoover85} and the pressure by the Hoover barostat as modified by Melchionna {\it et al.}~\cite{melchionna93}, with respective time constants $\tau_T=1$~ps and $\tau_P=2$~ps. We integrate the equations of motions using a time step of 0.5~fs and periodic boundary conditions. We cut off the short-range interactions at 10~{\AA} and use the Ewald method to deal with the electrostatic interactions. We perform the simulations using the {\footnotesize DL\_POLY 4} software~\cite{dlpoly}. We consider a system of the same size as in the FPMD simulations --- that is to say composed of 128 $\rm CO_3^{2-}$, 159 $\rm Li^+$ and 97 $\rm K^+$ --- to facilitate the comparison with FPMD data and favor longer simulation times over system size. This will be particularly relevant in the following for the runs needed to calculate the shear viscosity and the ionic conductivity. The starting randomized liquid configurations were generated by the {\footnotesize MOLDY} software~\cite{moldy}. Using our optimized force field we obtain $NPT$ densities of 1.96~g/cm$^3$, 1.90~g/cm$^3$ and 1.85~g/cm$^3$ and 900, 1000 and 1100~K respectively. This is in very good agreement with the experimental densities, and a significant improvement with respect to the Tissen and Janssen force field. As a matter of fact, for the sake of obtaining reasonable pressures, they had to reduce the density of the simulation box by 10\% with respect to the experimental one for the $\rm Li_2CO_3-K_2CO_3$ (62:38 mol\%) eutectic mixture~\cite{tissen94}.

\begin{figure}[hp]
\includegraphics[scale=0.7]{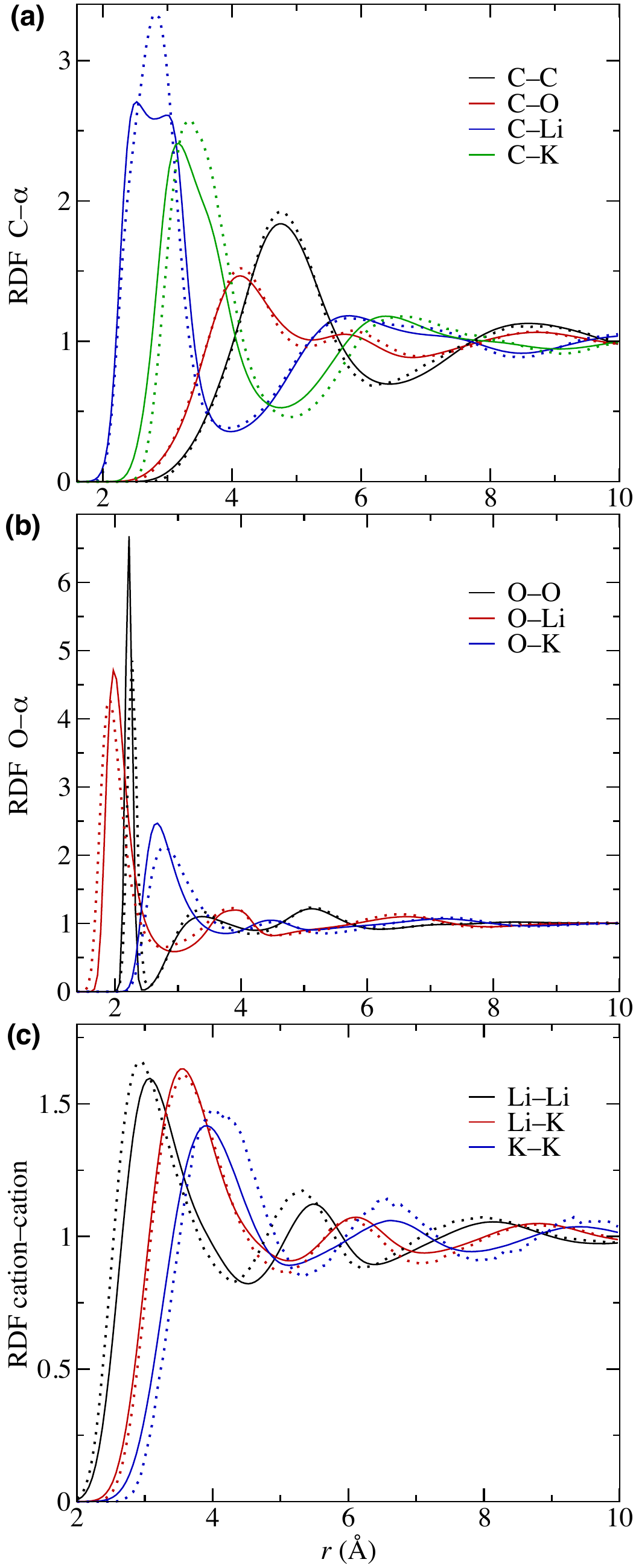}
\caption{Comparison of the RDFs obtained at $T=1000$~K with our optimized force field (solid lines) and by using FPMD simulations (dashed lines). In (a) we plot the C--C, C--O, C--Li and C--K RDFs. For C--O only the intermolecular part is shown. In (b) we show the O--O, O--Li and O--K RDFs. Panel (c) displays the cation--cation RDFs, Li--Li, Li--K and K--K.}
\label{fig:1}
\end{figure}

In order to make sure that the classical force field reproduces as well as possible the microscopic structure of the melt, we compute the RDFs from constant-volume $NVT$ simulations of 10~ns at the target temperatures, using starting configurations with the correct experimental density. We compare the RDFs obtained by the optimized classical model to the distributions functions obtained by FPMD simulations. We consider here in detail the case for $T=1000$~K; at the other two temperatures investigated we observe analogous trends. The comparison between classical and first principles RDFs is shown in Fig.~\ref{fig:1}(a) for the $\rm C-\alpha$ pairs, $\alpha=\rm C,O,Li,K$, in Fig.~\ref{fig:1}(b) for $\rm O-\alpha$ pairs, $\alpha=\rm O,Li,K$, and in Fig.~\ref{fig:1}(c) for cation--cation pairs. In Fig.~\ref{fig:1}(a), we observe how the agreement between the positions and the intensities of the peaks of the classical and first-principles C--C and C--O RDFs is remarkably good. A slight shift to longer distances is observed in classical simulations for the C--K pair. In the C--Li case, while the positions of the peaks are precisely reproduced by the classical model, the first peak appears split in the classical case. However, even in the first-principles case a shoulder in the peak can be observed at approximately {2.5~\AA}, corresponding to the first sub-peak of the classical case. Furthermore, the coordination number obtained by integrating the C--Li RDF until the first minimum occurring at $r_{\rm min}=3.975$~{\AA} is practically identical in the classical ($n_{\rm C-Li}=4.74$) and in the first-principles case ($n_{\rm C-Li}=4.73$). Fig.~\ref{fig:1}(b) shows that the agreement in the $\rm O-\alpha$ RDFs is quite good with minor shifts of the positions of the peaks to longer/shorter distances in the O--Li/O--K classical RDF. We consider the cation--cation case in Fig.~\ref{fig:1}(c). Despite the slight shift towards longer/shorter distances in the classical Li--Li/K--K classical RDFs, we see that the Li--K classical RDF reproduces well the data obtained from first-principles simulations.  The overall agreement between the classical and FPMD appears satisfactory, validating the force fields parameters optimized in this work.

\subsection{Classical force field simulations}

Provided with the force field derived as described in Sec.~\ref{sec:ff}, we then set out to perform classical MD simulations at different pressures, from atmospheric pressure to 5~GPa, and at the target temperatures, $T=900$~K, 1000~K and 1100~K. We first run the system in the $NPT$ ensemble for 0.5~ns. The simulation details are as in Section~\ref{sec:ffvalid}. After constant--pressure equilibration, production runs for the calculation of dynamical quantities are performed in the microcanonical $NVE$ ensemble.

The diffusion coefficients are calculated from 0.5~ns $NVE$ runs in which the trajectory is stored every 20~fs. The diffusion coefficient can be calculated from the simulation trajectory via the mean square displacement and the Einstein relation:
\begin{equation}
D_k= \frac{1}{N_k} \lim_{t\rightarrow\infty} \frac{1}{6t}\sum_{i\in k} \left \langle | \mathbf{r}_i(t)-\mathbf{r}_i(0)|^2\right \rangle
\end{equation}
where $N_k$ is the number of ions of kind $k$ ($k=\rm CO_3^{2-}, Li^+, K^+$), and $\langle\cdots\rangle$ represents an average performed over time origins. The viscosity is computed from $NVE$ runs of variable length, from 5 to 15~ns, according to the rapidity of convergence at the different thermodynamic conditions. We calculate the viscosity using the Green--Kubo relation:
\begin{equation}\label{eq:visc}
\eta = \frac{V}{k_B T} \int_0^\infty \left\langle \Pi_{\alpha\beta}(t)\Pi_{\alpha\beta}(0)\right\rangle \, {\rm d}t
\end{equation}
where $\Pi_{\alpha\beta}$ is any off-diagonal element of the pressure tensor $\mathbf\Pi$. In practice, in order to improve the statistics, an average is performed over the five independent off-diagonal terms of the pressure tensor $\Pi_{xy}$, $\Pi_{xz}$, $\Pi_{yz}$, $(\Pi_{xx}-\Pi_{yy})$ and $(2\Pi_{zz}-\Pi_{xx}-\Pi_{yy})$. We accumulate the values of the pressure tensor every 20~fs. Finally, we calculate the ionic conductivity from 15~ns $NVE$ runs in which the total dipole moment is sampled every 10~fs. The ionic conductivity is then calculated as:
\begin{equation}\label{eq:conduc}
\sigma=\frac{1}{k_BTV}\lim_{t\rightarrow\infty} \frac{1}{6t}\left \langle \left | \sum_i q_i (\mathbf{r}_i(t)-\mathbf{r}_i(0)) \right |^2\right\rangle
\end{equation}
where $q_i$ is the electrical charge of particle $i$. The expression within the angular brackets can also be written as $\left|\mathbf{M}(t) - \mathbf{M}(0)\right|^2$, where $\mathbf{M}(t) = \sum q_i \mathbf{r}_i(t)$ is the macroscopic dipole moment of the sample. We can notice that for the ionic conductivity, the time correlation function to be accumulated involves a collective quantity, which, in general, has a greater statistical noise than a {\it self} quantity such as the mean square displacement. This statistical noise can be reduced performing longer simulation runs, and this is the reason why we use much longer runs for the ionic conductivity with respect to the calculation of the diffusion coefficients.

\section{Results}

The results obtained from classical MD simulations of the $\rm Li_2CO_3-K_2CO_3$ (62:38 mol\%) eutectic mixture using our optimized force field are presented below. In Section~\ref{sec:eos} we present the thermodynamics of the eutectic mixture, and in particular its equation of state. In Section~\ref{sec:struct} we investigate the evolution of the structure as function of pressure. Finally, in Section~\ref{sec:dynamics} we focus on the dynamical properties: diffusion coefficients, viscosity and ionic conductivity.

\begin{figure}[!h]
\includegraphics[width=8cm]{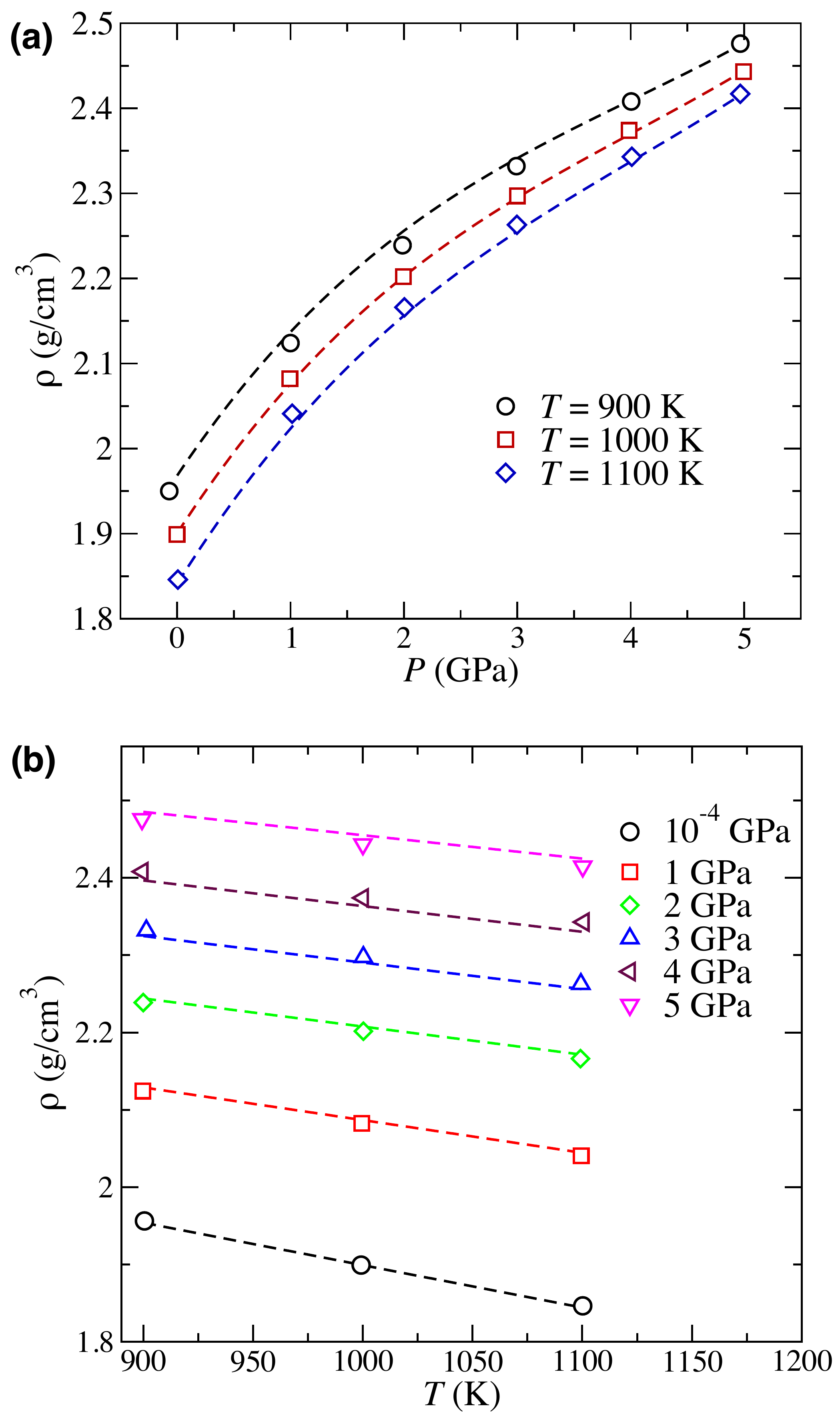}
\caption{Equation of state of the eutectic mixture: (a) isothermal EOS $\rho(P)$; (b) isobaric EOS $\rho(T)$. The symbols represent values calculated in the MD simulations, the dashed lines are EOS fits as described in the text.}
\label{fig:2}
\end{figure}

\subsection{Equation of State}\label{sec:eos}

We study the behavior of the {\it equation of state} (EOS) based on $NPT$ simulations at various values of temperature and pressure. Fig.~\ref{fig:2}(a) shows the isothermal EOS $\rho_T(P)$, while Fig.~\ref{fig:2}(b) shows the isobaric EOS $\rho_P(T)$ EOS. In order to be able to predict the density as a function of pressure at constant temperature, we attempt to fit the points shown in Fig.~\ref{fig:2} with an empirical relation. We find that the formula:
\begin{equation}
\rho_T(P)=AT^{-1/3} +BT^{1/3}P +CP^2+DP^3
\end{equation}
provides a reasonable approximation of the isothermal EOS shown in Fig.~\ref{fig:2}, with the following values for the parameters: $A = 19.033\, \rm g\, K^{1/3}cm^{-3}$; $B=2.063\ 10^{-2}\, \rm g\, K^{-1/3} GPa^{-1} cm^{-3}$; $C=-3.314\ 10^{-2}\, \rm g\, GPa^{-2} cm^{-3}$; and $D=2.727\ 10^{-3}\, \rm g\, GPa^{-3} cm^{-3}$. Similarly, we find an empirical fit for the isobaric EOS:
\begin{equation}
\begin{cases}
	\rho_{P}(T) = \alpha(P) + \beta(P)T\\
	\alpha(P)=2.4474 + 6.1992\ 10^{-2}\ P\\
	\beta(P)=-5.4831\ 10^{-4} + 1.6840\ 10^{-4} P - 4.7540\ 10^{-5} P^2 + 4.7401\ 10^{-6} P^3
\end{cases}
\end{equation}
with $\rho$ in units of g/cm$^3$, $T$ in K and $P$ in GPa. These empirical EOS predict with reasonable accuracy the thermodynamics of the $\rm Li_2CO_3-K_2CO_3$ (62:38 mol\%) eutectic mixture within the temperature and pressure spans of $T=900$--1100~K and $P=0$--5~GPa.

\begin{figure}[!h]
\includegraphics[width=8cm]{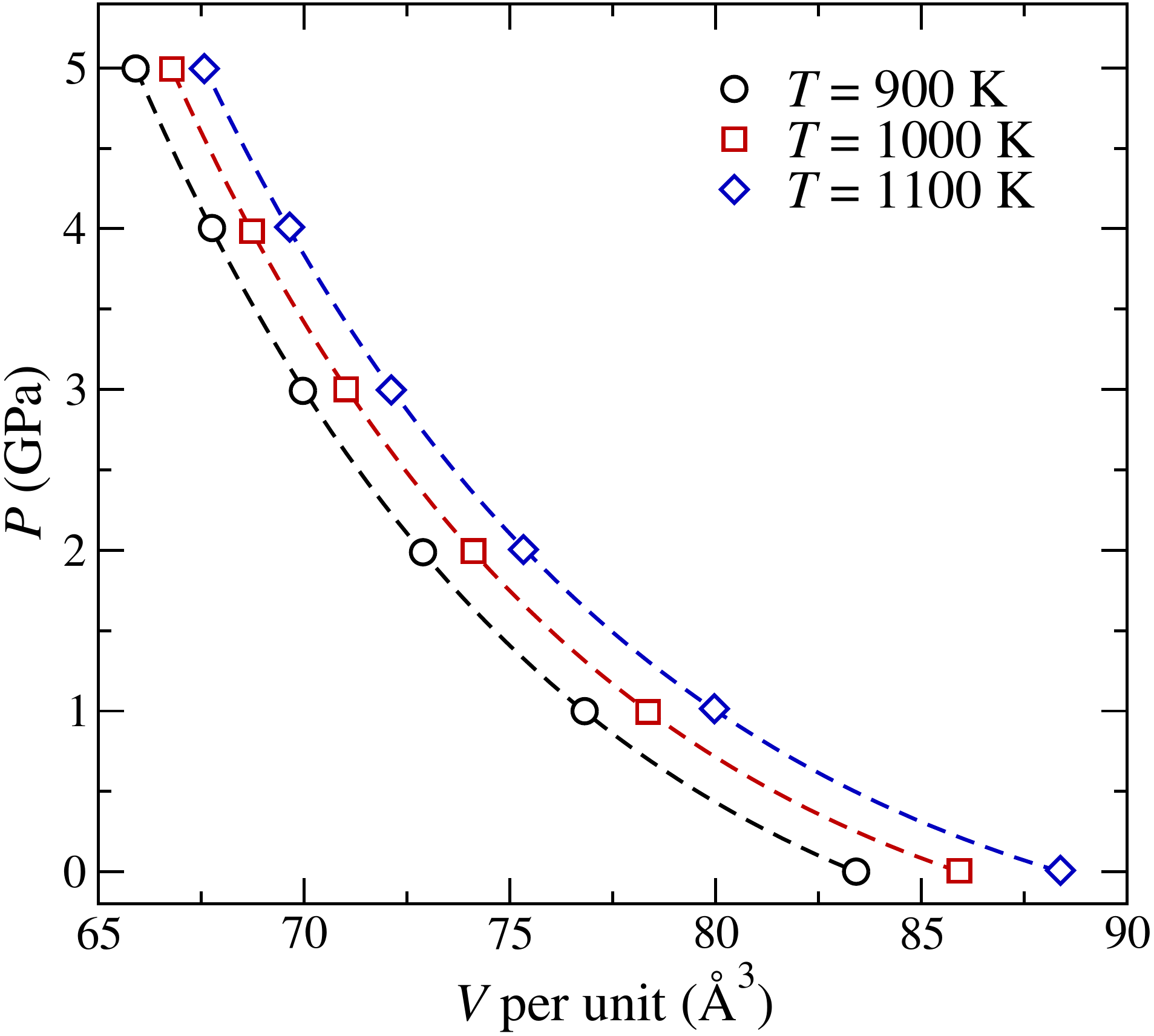}
\caption{Isothermal $P_T(V)$ equation of states. Symbols mark the values obtained in MD simulations, dashed lines are fits by third-order Birch-Murnaghan equations of state. The volume is shown per formula unit of carbonate. The value of the reference volume $V_0$ (see Eq.~\ref{eq:birch}) is 83.34~\AA$^3$/unit at $T=900$~K (i.e. $\rho_0 = 1.960$~g/cm$^3$), 85.93~\AA$^3$/unit at $T=1000$~K (i.e. $\rho_0 = 1.901$~g/cm$^3$) and 88.38~\AA$^3$/unit at $T=1100$~K (i.e. $\rho_0 = 1.848$~g/cm$^3$).}
\label{fig:3}
\end{figure}

To go beyond these empirical EOS, we also fit our $P_T(V)$ data using the conventional third-order Birch--Murnaghan EOS~\cite{birch47}:
\begin{equation}\label{eq:birch}
P_T(V)= \frac{3B_0}{2} \left [ \left( \frac{V_0}{V}\right)^{7/3} - \left( \frac{V_0}{V}\right)^{5/3} \right ] \left\{ 1+\frac{3}{4} (B'_0-4) \left [ \left( \frac{V_0}{V}\right)^{2/3} - 1 \right ] \right \}
\end{equation}
where $B_0$ is the bulk modulus, $B'_0$ is the derivative of the bulk modulus with respect to pressure and $V_0$ is the reference volume. The resulting fits are depicted in Fig.~\ref{fig:3}. We find that the data are perfectly described by the Birch--Murnaghan EOS, with the respective reference densities at $T=900/1000/1100$~K that are fully consistent with the densities at $P=0$. We obtain for the bulk modulus and its pressure derivative the values $B=8.74/7.56/6.81$~GPa and $8.01/8.31/8.06$ respectively at $T=900/1000/1100$~K. The reference densities vary linearly with temperature, $\rho_0(T) = 1.903 - 5.6\ 10^{-4} (T - 1000)$, with $T$ in K and $\rho_0$ in g/cm$^3$.

\subsection{Structure}\label{sec:struct}

\begin{figure}[!h]
\includegraphics[width=\textwidth]{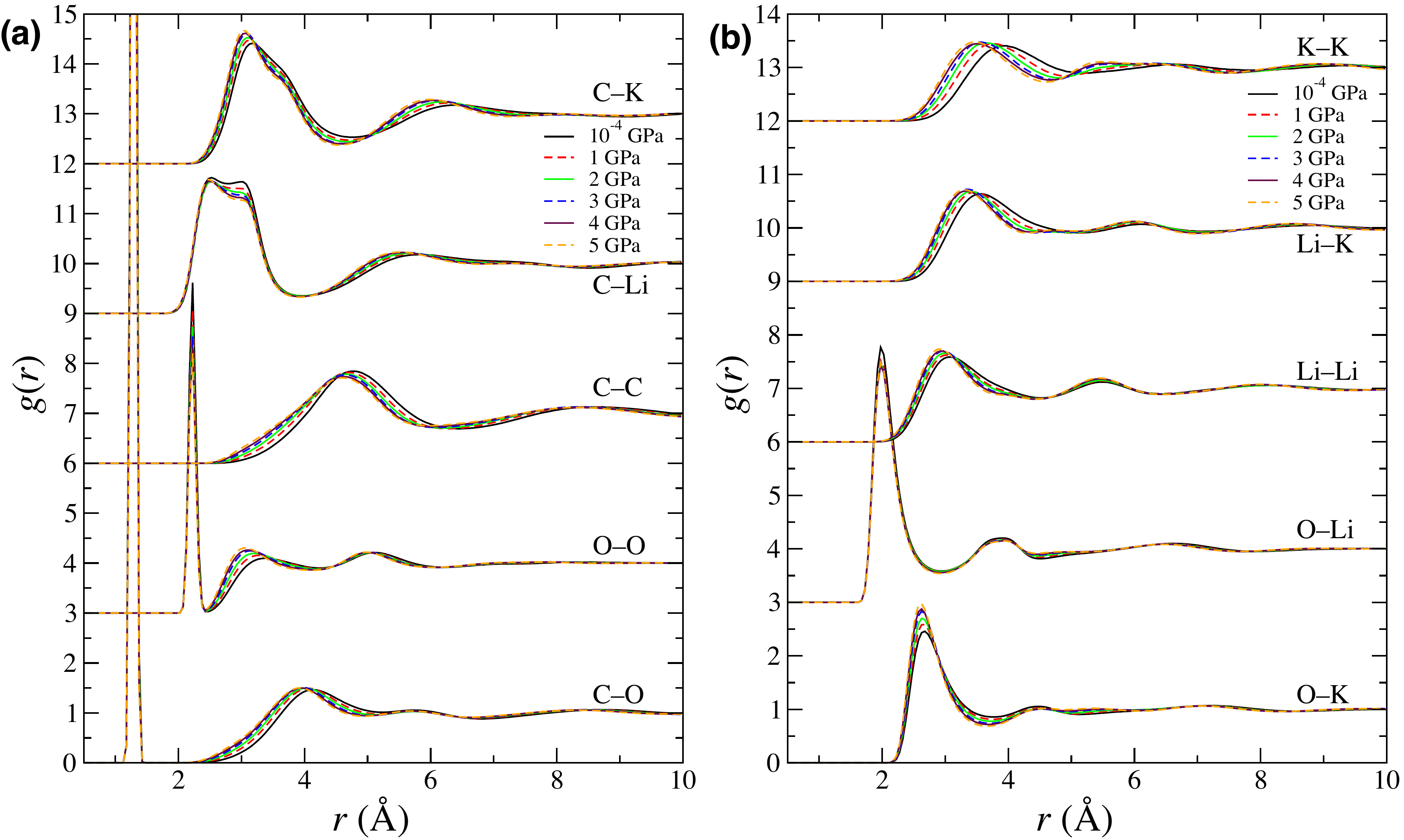}
\caption{RDFs at the different pressures investigated. (a) C--O, O--O, C--C, C--Li and C--K RDFs; (b) O--K, O--Li, Li--Li, Li--K and K--K RDFs. In both panels each RDF has been shifted on the vertical axis by $3n$, $n=0,1,2,3,4$ to facilitate the visualization.
}
\label{fig:4}
\end{figure}

Next, we focus on the evolution of the structural properties with pressure. We do so by looking, in Fig.~\ref{fig:4}, at all the RDFs at $T=1000$~K and at all simulated pressures, from 0 to 5 GPa. We select again the intermediate temperature, $T=1000$~K, since the changes in the RDFs with temperature at a given pressure are less significant than the modifications in the RDFs with pressure at a given temperature. We calculate the RDFs from the same 0.5~ns $NVE$ runs used to calculate the diffusion coefficients. We have seen before that the RDFs at ambient pressure reproduce satisfactorily the ones calculated in FPMD simulations. At our composition, the only RDFs previously reported, see Ref.~\cite{tissen94}, by classical MD, have been for C--Li, C--K, O--Li and O--K, calculated at $T=1200$~K using the Tissen and Janssen force field~\cite{tissen90}. Although minor differences can be observed between our RDFs and the ones calculated in Ref.~\cite{tissen94}, the main features, i.e. the positions of the peaks, are in good agreement.

Turning to the effect of pressure on the RDFs, we can observe in Fig.~\ref{fig:4} that for the C--O and O--O pairs the first peak (intramolecular distances) remains unaffected, while the position of the second peak moves to shorter distances upon increasing the pressure. Analogously, the first peak of the C--C RDF moves inwards when pressure increases, reflecting the overall change in density of the system. More subtle changes are observed when looking at the C--Li and C--K pairs. In the C--Li case, when the pressure increases, the first sub-peak at 2.5~\AA\ gets enhanced significantly while the intensity of the other one, at 3.0~\AA, significantly reduces. The peak position of the second shell also moves inwards upon increasing the pressure. The same occurs for the second shell of C--K. In this case, however, the first peak appears to sharpen at higher pressures and its position shifts to shorter distances, although a shoulder at about 3.55~\AA\ becomes increasingly visible at high pressures.

The O--Li RDF appears practically unchanged at different pressures. Conversely, an important effect is visible in the O--K structure. In this case, when the pressure increases, the height of the first peak increases and its position shifts to shorter distances. At the same time, the second shell becomes increasingly broader. The cation--cation RDFs undergo similar modifications upon increasing pressure. In fact for the Li--Li, Li--K, and K--K pairs we observe that for both the first and second peak there is an increase in the intensity and a shift to lower distances. In the Li--K case at high pressure, a small intermediate peak appears between the first and the second one. For the K--K pair the second shell becomes increasingly broad with pressure.

In summary, the main effect of increasing the pressure, apart from the overall compression which brings all pairs closer together, seems to be the shortening of the O--K distances. In other words, more K$^+$ ions are able to sit at short distances from the carbonate oxygen atoms. This in turn explains the change in the C--K first shell.

\begin{table}[htbp]
\caption{Self diffusion coefficients at ambient pressure: comparison between values calculated in this work to experimental data~\cite{janz82} and MD simulations by Tissen {\it et al.}~\cite{tissen94}, Koishi {\it et al.}~\cite{koishi00} and Costa {\it et al.}~\cite{costa08} using a non polarizable (NP) or polarizable (P) force field. Note that the temperatures are different. We report our values at $T=1100$~K to facilitate the comparison.}
\begin{center}
\begin{ruledtabular}
\begin{tabular}{lcccccc}
 & Exp. & Tissen & Koishi & Costa NP & Costa P & This work\\
$T$~(K) & 1100 & 1200 & 1200 & 1073 & 1073 & 1100\\
\hline
$D_{\rm CO_3^{2-}}$ ($10^{-5}$ cm$^2$/s) & 3.52 & 1.9 & 0.75 & 0.55 & 0.94 & 1.63\\
$D_{\rm Li^{+}}$ ($10^{-5}$ cm$^2$/s) & -- & 4.3 & 2.86 & 2.55 & 4.42 & 5.38\\
$D_{\rm K^{+}}$ ($10^{-5}$ cm$^2$/s)& 3.81 & 4.6 & 1.55 & 1.65 & 2.61 & 4.99
\end{tabular}
\end{ruledtabular}
\end{center}
\label{tab:2}
\end{table}

\subsection{Dynamics}\label{sec:dynamics}

We now move to the investigation of the dynamical properties of our systems. We will discuss in particular the behavior of the diffusion coefficients, the viscosity and the ionic conductivity. We begin our discussion from the diffusion coefficients of the three molecular species present, $D_{\rm CO_3^{2-}}$, $D_{\rm Li^+}$ and $D_{\rm K^+}$. First, we compare the values of the diffusion coefficients obtained in our model to available experimental data and previous MD simulations results, see Table~\ref{tab:2}. We consider here the values of the diffusion coefficients at $T=1100$~K since the previously available simulation data are either at $T=1073$~K or at $T=1200$~K.

Experimentally the only data available are for $\rm CO_3^{2-}$ and $\rm K^+$ at ambient pressure and can be extracted from the Arrhenius relations $D=D_\infty\exp(-E_A/RT)$ as reported in Ref.~\cite{janz82} in the temperature range from 890~K to 1135~K. For $\rm CO_3^{2-}$, $D_\infty=794\ 10^{-5}\ \rm cm^2/s$ and $E_A=49.62$~kJ/mol, for $\rm K^+$, $D_\infty=726\ 10^{-5}\ \rm cm^2/s$ and $E_A=48.07$~kJ/mol. Using these relations, we can calculate the expected experimental value at $T=1100$~K as shown in Table~\ref{tab:2}. Tissen {\it et al.}~\cite{tissen94} and Koishi~{\it et al.}~\cite{koishi00} have both conducted 40~ps MD simulations at $T=1200$~K, yet the data from Koishi~{\it et al.} indicate much lower diffusion coefficients. This may be related to the fact that Tissen~{\it et al.} reduced the density of their sample by 10\% in order to adjust the pressure. Costa and Ribeiro conducted 80~ps runs at $T=1073$~K and $T=1200$~K using the Tissen and Janssen force field, as well as a polarizable force field. Comparing with these results from the literature, our force field shows a significantly improved diffusivity for carbonate with respect to the Tissen and Janssen force field, and performs even better with respect to Costa's polarizable model. For the diffusion coefficient of  $\rm K^+$ ions, the deviation from the reported experimental value is of the same magnitude as for Costa's polarizable model, but while the latter underestimates it, our force field appears to overestimate it. The Tissen and Janssen force field produces a $D_{\rm K^+}$ even lower than Costa's polarizable force field.

\begin{figure}[p]
\includegraphics[width=72mm]{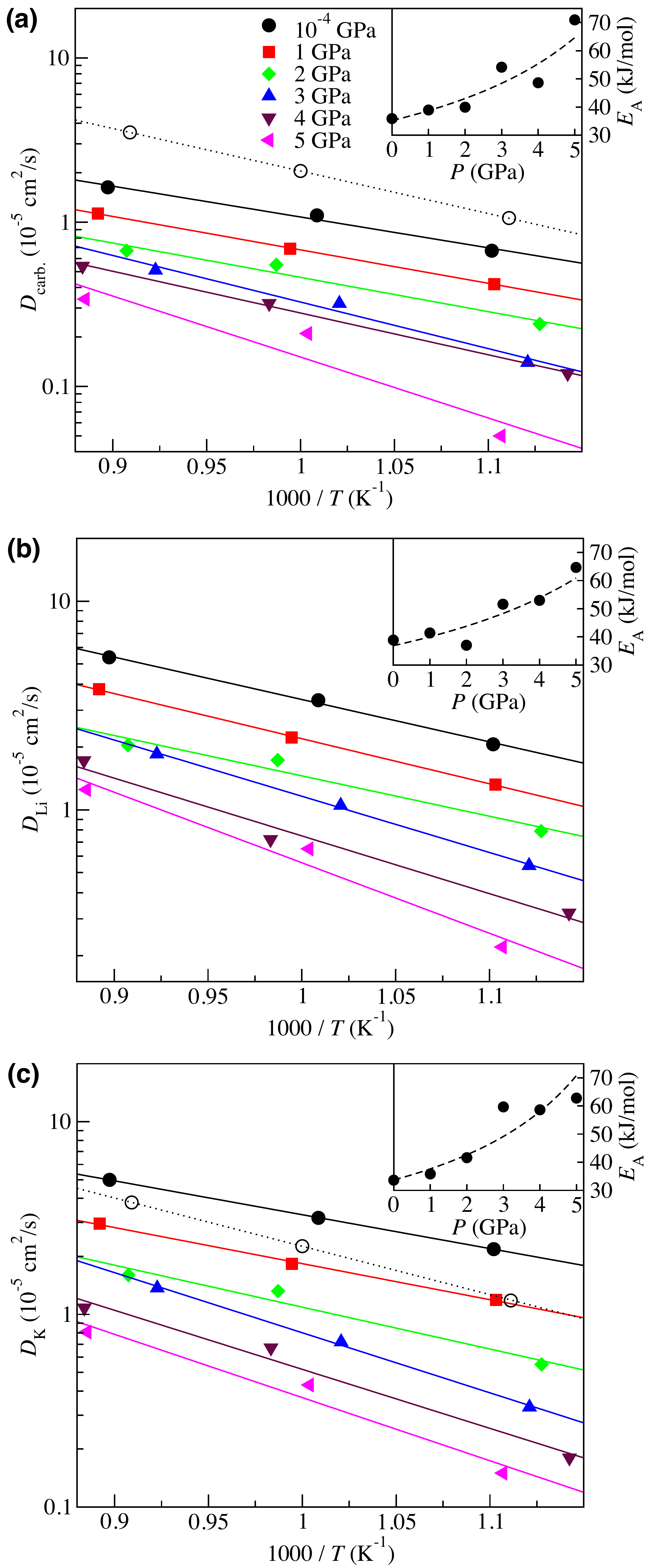}
\caption{Ionic self diffusion coefficient (in logarithmic scale) as a function of inverse temperature and pressure for (a) carbonate ions, (b) $\rm Li^+$ and (c) $\rm K^+$. Symbols are simulations results, lines are fits to Arrhenius' law, $D=D_\infty \exp(-E_A/RT)$. The parameters obtained for each species and pressure are reported in Table~\ref{tab:app1}. The open circles and the dotted lines in panel (a) and (c) are the experimental values reported in Ref.~\cite{janz82}. In insets we show the evolution of $E_A$ with pressure, fitted using the function $E_A(P)=(A-BP)^{-1}$. The parameters obtained are (a) $A=0.0284$~mol/kJ and $B=0.0026$~mol/(kJ GPa) for $\rm CO_3^{2-}$; (b) $A=0.0271$~mol/kJ and $B=0.0021$~mol/(kJ GPa) for $\rm Li^+$; (c) $A=0.0297$~mol/kJ and $B=0.0031$~mol/(kJ GPa) for $\rm K^+$.}
\label{fig:5}
\end{figure}

Therefore, while our force field cannot exactly match the experimental values, we see that it is of the same accuracy of Costa's polarizable model in the reproduction of $D_{\rm K^+}$ and it is significantly better for the diffusivity of carbonate. In Fig.~\ref{fig:5} we plot the temperature and pressure dependence of the self diffusion coefficients $D_{\rm CO_3^{2-}}$ (panel a), $D_{\rm Li^{+}}$ (panel b) and $D_{\rm K^{+}}$ (panel C). We fit the simulation data with Arrhenius' law $D=D_\infty\exp(-E_A/RT)$; the values of the parameters $D_\infty$ and $E_A$ extracted by this procedure are tabulated in Table~\ref{tab:app1} in the Appendix. We compare the points at the three temperatures investigated, $T=900$~K, $T=1000$~K and $T=1100$~K and the Arrhenius curves to the experimental behavior as reported in Ref.~\cite{janz82}. We can observe how the agreement with experimental data tends to get better at lower temperatures for carbonate ions. For $\rm K^+$ instead, the (upward) shift with respect to the experimental values increases at lower temperatures. In all cases the activation energy $E_A$ appears lower in our simulations than the experimental values. For comparison, the activation energy found here for  $\rm Li_2/K_2CO_3$ (62:38 mol\%) are of the same order of magnitude as that for $\rm CaCO_3$ obtained from first-principle simulations in Ref.~\cite{vuilleumier2014}. In the insets of Fig.~\ref{fig:5} we plot the activation energy $E_A$ as a function of pressure. We see that in all cases $E_A$ tends to increase with pressure, a behavior that is qualitatively in agreement with the experimental results, although we remark that our results numerically deviate from the experimental data, mainly due to the lower/higher (carbonate/K$^+$) values of the diffusion coefficients. This increase in the activation energy as a function of pressure is characterized by an activation volume $V_A$ through $\frac{\partial E_A(P)}{\partial P}=V_A(P)$. We have fitted the simulation results by an empirical relation $E_A(P)=(A-BP)^{-1}$ to extract the activation volumes. From ambient pressure to 5~GPa, the activation volumes range from 3.2~cm$^3$/mol to 8.0~cm$^3$/mol for $\rm CO_3^{2-}$, from 2.9~cm$^3$/mol to 7.6~cm$^3$/mol for $\rm Li^+$ and from 3.5~cm$^3$/mol to 15.3~cm$^3$/mol for $\rm K^+$. We oberve that $\rm Li^+$ has the smallest activation volume while $\rm K^+$ the largest. These values are notably higher than those found for $\rm CaCO_3$ at high pressures from first-principle simulations in Ref.~\cite{vuilleumier2014}. This may be due to the use of a non-polarizable force-field.

\begin{table}[htbp]
\caption{Comparison of the experimental values of the viscosity at ambient pressure for $\rm Li_2CO_3-K_2CO_3$ (62:38 mol\%) at $T=905$~K, $T=993$~K and $T=1079$~K from Ref.~\cite{kanai2013}, those of the (60:40 mol\%) mixture at  $T=1000$~K and $T=1100$~K~\cite{janz79}, and the values obtained for $\rm Li_2/K_2CO_3$ (62:38 mol\%) in this work by MD simulations.}
\begin{center}
\begin{ruledtabular}
\begin{tabular}{lcc}
 & Exp. & This work\\ \hline
$\eta$ (mPa $\cdot$ s) at $T=900$~K & 6.99~\cite{kanai2013} & 6.31\\
$\eta$ (mPa $\cdot$ s) at $T=1000$~K & 4.81~\cite{kanai2013} -- 5.10~\cite{janz79} & 4.19\\
$\eta$ (mPa $\cdot$ s) at $T=1100$~K & 3.54~\cite{kanai2013} -- 3.51~\cite{janz79} & 3.04
\end{tabular}
\end{ruledtabular}
\end{center}
\label{tab:3}
\end{table}

We now move to the discussion of another transport property, the shear viscosity $\eta$. This property is much harder to obtain from MD simulations than diffusion coefficients, due to the long runs necessary to converge the autocorrelation function of the pressure tensor (see Eq.~\ref{eq:visc}). As a matter of fact, we have not found any previous calculations of the viscosity for the eutectic $\rm Li_2CO_3-K_2CO_3$ mixture, and ours is the first attempt to calculate the viscosity of this system by MD simulations. In Table~\ref{tab:3}, we compare the values of viscosity at ambient pressure, calculated in our simulations, to the experimental values from Ref.~\cite{kanai2013} for the same mixture at $T=905$~K, $T=993$~K and $T=1079$~K, and to the values from Ref.~\cite{janz79} for $\rm Li_2CO_3-K_2CO_3$ mixture with composition of (60:40 mol\%). We can see that despite the slight difference in composition, the values obtained in our simulations are not so far from the experimental ones: our results are within 20\% of the experimental viscosity, a rather good result considering that we are using here a simple non--polarizable force field.

\begin{figure}[p]
\includegraphics[width=8cm]{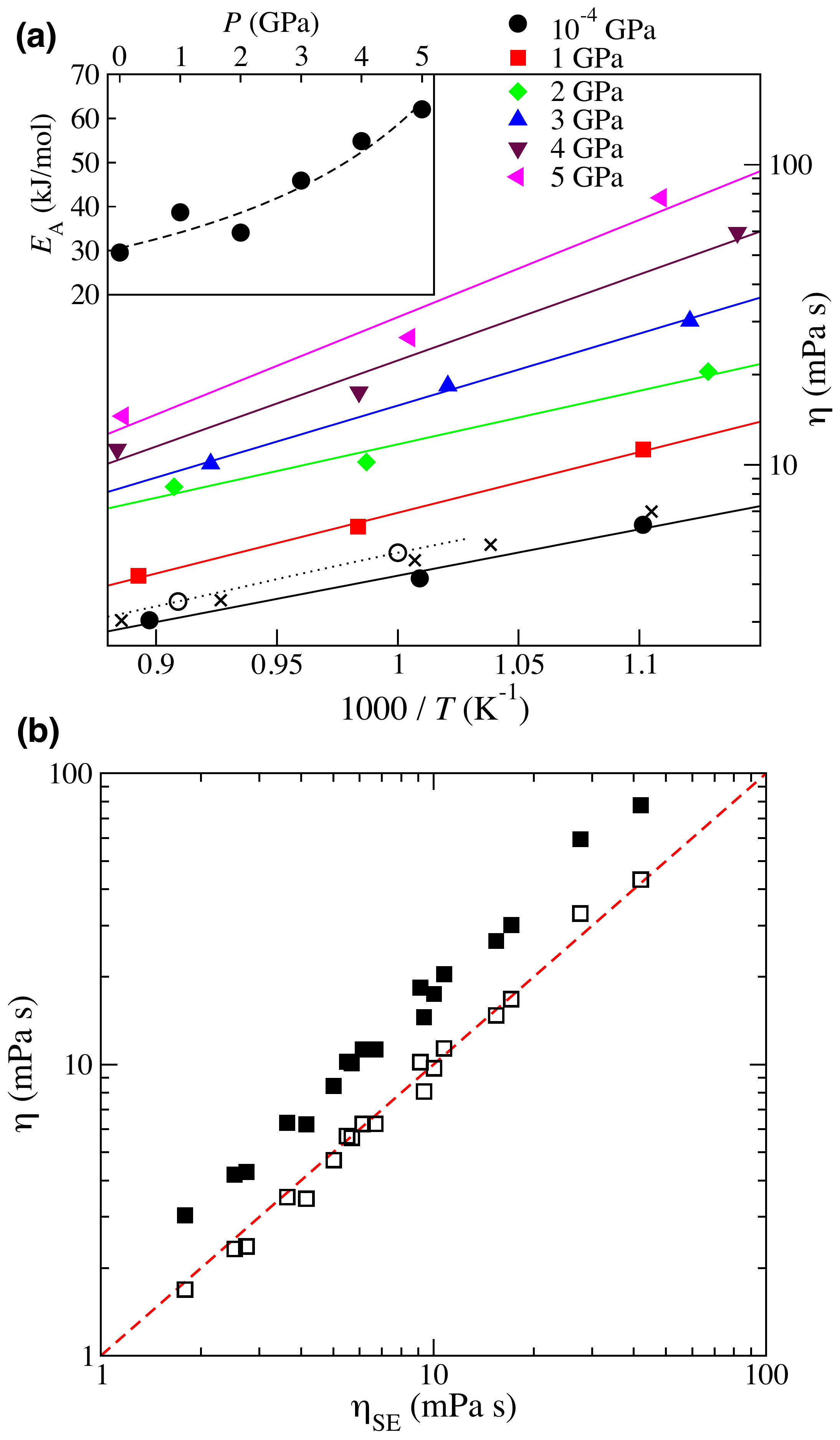}
\caption{(a) Shear viscosity $\eta$ (in logarithmic scale) as a function of inverse temperature. Filled symbols correspond to MD simulations, lines represent Arrhenius fits $\eta=\eta_\infty \exp(E_A/RT)$. The parameters $\eta_\infty$ and $E_A$ extracted from the fits are reported in Table~\ref{tab:app2}. $\times$ symbols are experimental data from Ref.~\cite{kanai2013}. The open circles and the dotted line are experimental values reported in Ref.~\cite{janz79}. In inset we show how $E_A$ varies with pressure, including a fit of the form $E_A(P)=(A-BP)^{-1}$, with $A=0.0327$~mol/kJ and $B=0.0034$~mol/(kJ GPa). (b) Comparison in a log--log plot between the values of the viscosity calculated from the simulations using the Green--Kubo relation (Eq.~\ref{eq:visc}) and deduced from the Stokes--Einstein relation (filled squares). The red dashed line corresponds to $\eta=\eta_{\rm SE}$, the open squares are obtained by setting $\eta'=\eta/1.8$.}
\label{fig:6}
\end{figure}

In Fig.~\ref{fig:6}(a), we plot the viscosity as a function of inverse temperature at the different pressures investigated. We fit our simulation data points to an Arrhenius law $\eta=\eta_\infty\exp(E_A/RT)$, whose fitted parameters $\eta_\infty$ and $E_A$ are reported in Table~\ref{tab:app2} in the Appendix. In the same figure we also plot the experimental curve for the viscosity of $\rm Li_2CO_3-K_2CO_3$ (60:40 mol\%) from Ref.~\cite{janz79} and the experimental points for the viscosity of $\rm Li_2CO_3-K_2CO_3$ (62:38 mol\%) from Ref.~\cite{kanai2013}. We again observe a close agreement between the experimental and the simulated curves for $\eta_P(T)$. In the inset of Fig.~\ref{fig:6}(b), we plot the pressure dependence of the activation energy extracted from the Arrhenius fits described above. As in the case of the self diffusion coefficients, the activation energy increases with increasing pressure.

Finally, in Fig.~\ref{fig:6}(b) we compare the values of the viscosity obtained in our simulations to the values that would be predicted by the Stokes--Einstein relation:
\begin{equation}\label{eq:SE}
\eta=\frac{k_BT}{2\pi Dd}
\end{equation}
where in our case we define, following Ref.~\cite{vuilleumier2014}, $D$ as the weighted average of the ionic self diffusion coefficients: $D= (\sum_k N_k D_k)/ \sum_k N_k$, with $k= \rm CO_3^{2-}, Li^+, K^+$. For the hydrodynamic radius $d$, we adopt the following combination rule: $d=(d_{\rm C-C} +2d_{\rm C-Li}+2d_{\rm C-K}+d_{\rm Li-Li}+d_{\rm K-K})/7$, where $d_{\alpha-\beta}$ is the distance corresponding to the first peak of the pair RDF $g_{\alpha-\beta} (r)$ (see Fig.~\ref{fig:4}). We find that the viscosity predicted from the Stokes--Einstein relation with the assumptions just described and by inputing the values of the self diffusion coefficients as calculated from our simulations (see Fig.~\ref{fig:5}) is always lower than the ``true'' viscosity calculated directly from the Green--Kubo relation (Eq.~\ref{eq:visc}). In particular, we see that the Stokes--Einstein formula underestimates the viscosity by an approximately constant factor of about 1.8, see Fig.~\ref{fig:6}(b). If we rescale the calculated viscosity by this constant, it collapses on the Stokes--Einstein viscosity.

\begin{table}[htbp]
\caption{Comparison of the experimental values of the ionic conductivity at ambient pressure for $\rm Li_2CO_3-K_2CO_3$ (62:38 mol\%) at $T=923$~K and $T=973$~K from Ref.~\cite{lair2012}, those for a 60:40 mol\% mixture at $T=1000$~K and $T=1100$~K from Ref.~\cite{janz79}, and the values calculated in this work by MD simulations for the 62:38 mol\% mixture.}
\begin{center}
\begin{ruledtabular}
\begin{tabular}{lcc}
 & Exp. & This work\\\hline
$\sigma$ (S/m) at $T=900$~K &  113.4~\cite{lair2012} & 137.3\\
$\sigma$ (S/m) at $T=1000$~K & 155~\cite{lair2012} -- 197.5~\cite{janz79} & 210.7\\
$\sigma$ (S/m) at $T=1100$~K & 239.7~\cite{janz79} & 228.4\\
\end{tabular}
\end{ruledtabular}
\end{center}
\label{tab:4}
\end{table}

As a last quantity of interest, we consider now the ionic conductivity $\sigma$. As mentioned earlier in the Methods section, this quantity is more affected by statistical noise than the self diffusion coefficients, since it is derived from the autocorrelation function of a collective quantity, namely the total dipole moment. Despite the long simulation runs performed, the results for the ionic conductivity have higher statistical uncertainty than the ones for the diffusion coefficients. Nonetheless, we can appreciate in Table~\ref{tab:4} that the ionic conductivity values calculated with our force field are not so far from the experiments. We compare our values to the experimental ones available for the $\rm Li_2CO_3-K_2CO_3$ eutectic mixture from Ref.~\cite{lair2012} at $T=923$~K and $T=973$~K and to the ones for the 60:40 mol\% mixture at $T=1000$~K and at $T=1100$~K from Ref.~\cite{janz79}. The values calculated from the simulations differ from the experimental values at maximum by 21\%, which, again, is rather good for a non--polarizable force field.

The calculation of the ionic conductivity has been previously attempted by MD simulations. Koishi {\it et al.}~\cite{koishi00}, using the Tissen and Janssen potential obtained $\sigma=333$~S/m at $T=1200$~K by non--equilibrium MD simulations. Costa {\it et al.}~\cite{costa08}, using a Green--Kubo approach, obtained $\sigma=134$ and 145~S/m at $T=1073$~K and $\sigma=182$ and 237~S/m at $T=1200$~K, respectively for the Tissen and Janssen force field and the polarizable force field. Therefore, we see that our force field appears to behave better than the Tissen and Janssen force field in the reproduction of the ionic conductivity, and even better than Costa's polarizable model.

\begin{figure}[!h]
\includegraphics[width=80mm]{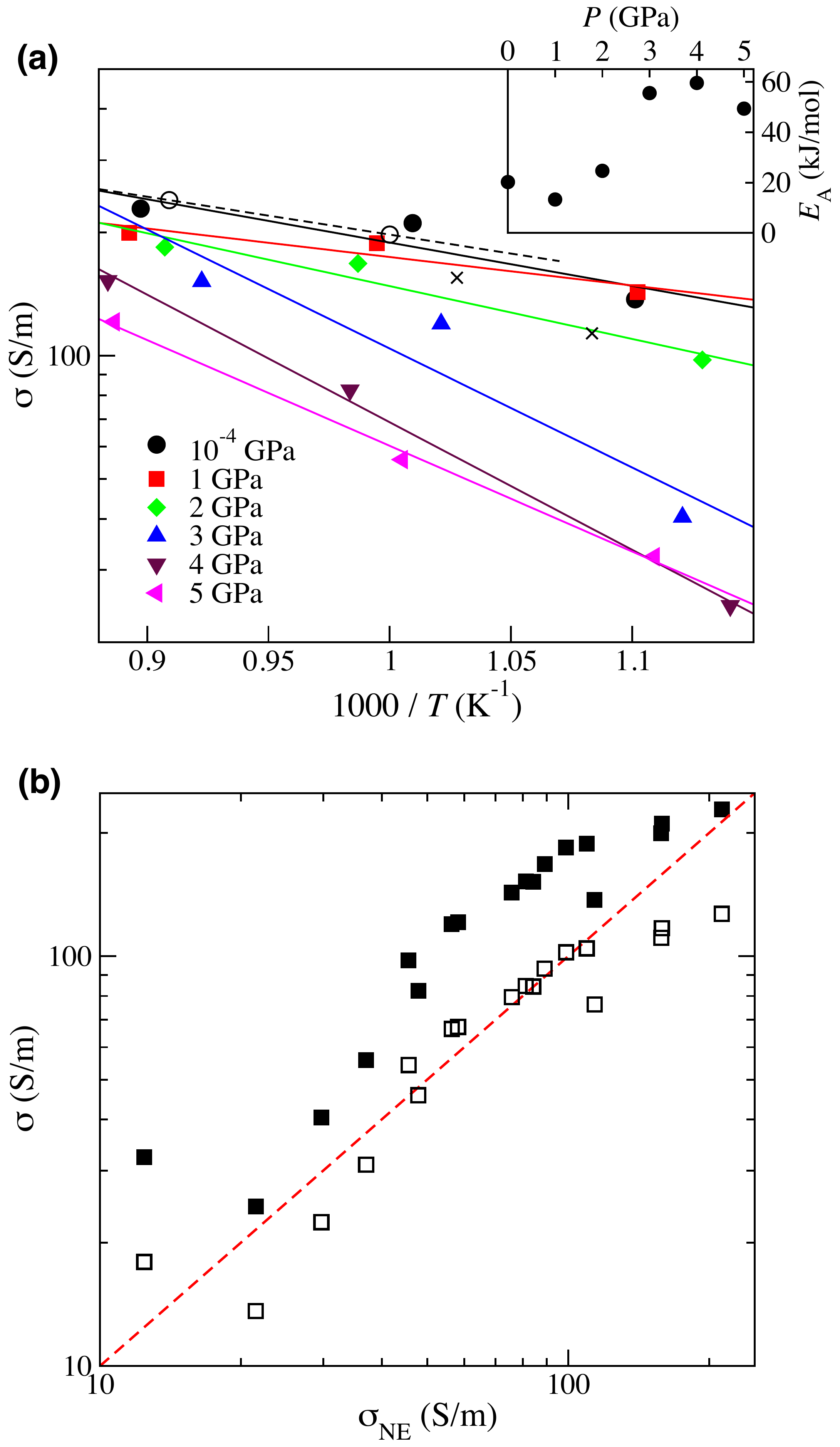}
\caption{(a) Ionic conductivity $\sigma$ (in logarithmic scale) calculated from MD simulations as a function of inverse temperature, at the different pressures investigated. Filled symbols correspond to MD simulations, lines represent Arrhenius fits $\sigma=\sigma_\infty \exp (-E_A/k_BT)$. The parameters $\sigma_\infty$ and $E_A$ extracted from the fits are given in Table~\ref{tab:app3}. The open circles and the dashed line correspond to the experimental data from Ref.~\cite{janz79}, while the $\times$ symbols correspond to the experimental data from Ref.~\cite{lair2012}. In the inset we show the activation energy $E_A$ as a function of pressure. (b) Log--log plot of the calculated conductivity $\sigma$ versus the Nernst--Einstein conductivity $\sigma_{\rm NE}$ (filled squares). The red dashes line corresponds to $\sigma=\sigma_{\rm NE}$ and the open squares are obtained by $\sigma'=\sigma_{NE} (1-\Delta)$, with $\Delta=-0.8$.}
\label{fig:7}
\end{figure}

In Fig.~\ref{fig:7}(a) we plot the ionic conductivity as a function of inverse temperature at all the investigated pressures. We can notice here that the statistical noise causes a somewhat less clear overall trend than for the viscosity or the diffusion coefficients. We fit the curves using Arrhenius relations $\sigma=\sigma_\infty \exp (-E_A/k_BT)$, with the values of parameters $\sigma_\infty$ and $E_A$ obtained listed in Table~\ref{tab:app3} in the Appendix. In the inset, we also plot the behavior of $E_A$ as a function of pressure. For the ionic conductivity, as opposed to the cases of the viscosity and of the self diffusion coefficient, we do not attempt to fit the data points due to statistical noise. However, we note that $E_A$ does, like for other quantities, increase with pressure.

In Fig.~\ref{fig:7}(b) we compare the values of the ionic conductivity calculated directly from the simulation trajectories using Eq.~\ref{eq:conduc} to the ones obtained by using the Nernst--Einstein (NE) relation:
\begin{equation}
\sigma = \frac{1}{k_BTV}\sum_k N_k q_k^2 D_k
\end{equation}
where $N_k$ is the number of ions of species $k$, $q_k$ their respective charge and $D_K$ their self diffusion coefficient. We note that we obtain values of the calculated conductivity that are systematically larger than the NE conductivity. Interestingly, we observe that also Costa~\cite{costa08} found a simulated value larger than the NE for the eutectic mixture at $T=1073$~K when using the Tissen and Janssen force field. Usually the real conductivity and the NE conductivity are related by $\sigma = \sigma_{NE} (1-\Delta)$, where the factor $(1-\Delta)$ is the inverse of the Haven ratio~\cite{murch82}. This factor is usually less than 1 and is interpreted to measure the degree of dissociation of the melt. In our case, as well as in Costa's, we can see by the ``rescaled'' points in Fig.~\ref{fig:7}(b) that the $1-\Delta$ factor is larger than 1: $\Delta\simeq -0.8$. It has been observed that in ionic liquid and molten salts, the increase in conductivity with respect to the the NE value may be due to the anticorrelation of the motion of opposite charges~\cite{kashyap11}.

\section{Conclusions}

We studied the molten carbonate $\rm Li_2CO_3-K_2CO_3$ eutectic mixture, with composition of 62:38 mol\%, in the 900--1100~K temperature range and at pressures up to 5~GPa. We optimized a non--polarizable empirical force field for this melt, using experimental and FPMD simulations data as reference. The force field derived in this work behaved reasonably well when compared to available experimental data, also considering its relative simplicity. Furthermore, this new force field was able to reproduce most properties better than previously available force fields. Using our newly developed force field, we characterized the thermodynamics, structure and dynamics of $\rm Li_2CO_3-K_2CO_3$. First, we calculated isothermal and isobaric EOS in the range of the thermodynamic conditions considered. We then assessed the microscopic structure of the melt. We showed the influence of pressure on the RDFs of all atom pairs, and noticed a specific shortening of the O--K distances at high pressure. We then studied dynamical properties, including self diffusion coefficients, viscosity and ionic conductivity. For each quantity, we reported both the temperature and pressure dependences. Activation energies were calculated by fits to Arrhenius equations, and in all cases the activation energies were observed to increase with pressure. Finally, we observed that ionic conductivity in the $\rm Li_2CO_3-K_2CO_3$ melt systematically exceeded the values expected from the Nernst--Einstein relation, a fact that might possibly be attributed to the anticorrelation of the motion of opposite charges. The microscopic mechanism at the origin of the deviations of the calculated viscosities and ionic conductivities from the Stokes-Einstein and the Nernst--Einstein relations, respectively, appears to be an interesting focus for future work.

\clearpage

\section*{Appendix A: Arrhenius fits' parameters}

\begin{table}[htbp]
\caption{Parameters of the Arrhenius fits $D=D_\infty \exp(-E_A/RT)$ shown in Fig.~\ref{fig:5} for each ionic species and pressure.}
\begin{center}
\begin{ruledtabular}
\begin{tabular}{lcc}
& $E_A (\rm kJ/mol)$ & $D_\infty$ ($10^{-5}\rm cm^2/s$)\\
\hline
$\rm CO_3^{2-}$\\
$P=P_{\rm atm}$& 35.88 & 80.70\\
$P=1$~GPa& 38.93 &73.45\\
$P=2$~GPa& 39.92 &56.31\\
$P=3$~GPa& 54.14 &222.63 \\
$P=4$~GPa& 48.61 &97.19\\
$P=5$~GPa& 70.94 &770.78\\
\hline
$\rm Li^{+}$\\
$P=P_{\rm atm}$& 38.87 & 362.89\\
$P=1$~GPa& 41.38 &318.56\\
$P=2$~GPa& 37.01 &124.97 \\
$P=3$~GPa& 51.60 &577.84\\
$P=4$~GPa& 52.97 &439.62\\
$P=5$~GPa& 64.67 & 1337.02\\
\hline
$\rm K^{+}$\\
$P=P_{\rm atm}$& 33.62 & 188.08\\
$P=1$~GPa& 35.82 &136.63\\
$P=2$~GPa& 41.61 &163.07\\
$P=3$~GPa& 59.66 &1052.89\\
$P=4$~GPa& 58.62 & 600.16\\
$P=5$~GPa& 62.78 & 706.48\\
\end{tabular}
\end{ruledtabular}
\end{center}
\label{tab:app1}
\end{table}

\begin{table}[htbp]
\caption{Parameters of the Arrhenius fits $\eta=\eta_\infty \exp(E_A/RT)$ shown in Fig.~\ref{fig:6} for each pressure.}
\begin{center}
\begin{ruledtabular}
\begin{tabular}{lcc}
& $E_A (\rm kJ/mol)$ & $\eta_\infty$ (mPa $\cdot$ s)\\
\hline
$P=P_{\rm atm}$& 29.58 & 0.12\\
$P=1$~GPa& 38.71 &0.07\\
$P=2$~GPa& 34.10 &0.19\\
$P=3$~GPa& 45.90 &0.06 \\
$P=4$~GPa& 54.84 &0.03\\
$P=5$~GPa& 62.07 &0.02\\
\end{tabular}
\end{ruledtabular}
\end{center}
\label{tab:app2}
\end{table}

\begin{table}[htbp]
\caption{Parameters of the Arrhenius fits $\sigma=\sigma_\infty \exp(-E_A/RT)$ shown in Fig.~\ref{fig:7} for each pressure.}
\begin{center}
\begin{ruledtabular}
\begin{tabular}{lcc}
& $E_A (\rm kJ/mol)$ & $\sigma_\infty$ (S/m)\\
\hline
$P=P_{\rm atm}$& 20.22 & 2152.53\\
$P=1$~GPa& 13.30 &862.30\\
$P=2$~GPa& 24.67 &2878.14\\
$P=3$~GPa& 55.54 &83283.02 \\
$P=4$~GPa& 59.58 &89411.09\\
$P=5$~GPa& 49.38 &22948.32\\
\end{tabular}
\end{ruledtabular}
\end{center}
\label{tab:app3}
\end{table}

\clearpage

\section*{Acknowledgements}

We thank Michel Cassir and Virginie Lair for fruitful discussions. We acknowledge funding by PSL Research University (project COOCAR, grant ANR-10-IDEX-0001-02) and Agence Nationale pour la Recherche (project ELECTROLITH, grant ANR-2010-BLAN-621-03). This work was performed using HPC resources from GENCI (grant  2015-082309).

\bibliographystyle{unsrt}

\begin{thebibliography}{99}

\bibitem{gaillard2008}
F. Gaillard, M. Malki, G. Iacono-Marziano, M. Pichavant and B. Scaillet,
Science
\textbf{322}, 1363--1365 (2008).

\bibitem{jones2013}
A. P. Jones, M. Genge and L. Carmody,
Rev. Mineral. Geochem.
\textbf{75}, 289--322 (2013).

\bibitem{li2011}
X. Li, N. Xu, L. Zhang and K. Huang,
ECS Trans.
\textbf{35}, 1267--1273 (2011).

\bibitem{vuilleumier2014}
R. Vuilleumier, A. Seitsonen, N. Sator and B. Guillot,
Geochim. Cosmochim. Ac.
\textbf{141}, 547 (2014).

\bibitem{corradini2015}
D. Corradini, F.-X. Coudert and R. Vuilleumier,
Nature Chem.
DOI: \href{http://dx.doi.org/10.1038/nchem.2450}{10.1038/nchem.2450}, in press (2016).

\bibitem{tissen90}
J. T. W. M. Tissen and G. J. M. Janssen,
Mol. Phys.
\textbf{71}, 413 (1990).

\bibitem{habasaki90}
J. Habasaki,
Mol. Phys.
\textbf{69}, 115 (1990).

\bibitem{costa08}
M. F. Costa and M. C. C. Ribeiro,
J. Mol. Liq.
\textbf{138}, 61 (2008).

\bibitem{cp2k} See \url{http://www.cp2k.org/}.

\bibitem{vandevondele2005a}
J. VandeVondele {\it et al.}
Comp. Phys. Commun.
\textbf{167}, 103 (2005).

\bibitem{lippert1997}
G. Lippert, J. Hutter and M. Parrinello,
Mol. Phys.
\textbf{92}, 477 (1997).

\bibitem{goedecker1996}
S. Goedecker, M. Teter and J. Hutter,
Phys. Rev. B
\textbf{54}, 1703 (1996).

\bibitem{hartwigsen1998}
C. Hartwigsen, S. Goedecker and J. Hutter,
Phys. Rev. B
\textbf{58}, 3641 (1998).

\bibitem{krack2005}
M. Krack,
Theor. Chem. Acc.
\textbf{114}, 145 (2005).

\bibitem{vandevondele2007}
J. VandeVondele and J. Hutter,
J. Chem. Phys.
\textbf{127}, 114105 (2007).

\bibitem{becke1988}
A. D. Becke,
Phys. Rev. A
\textbf{38}, 3098 (1988).

\bibitem{lee1988}
C. Lee, W. Yang and R. G. Parr,
Phys. Rev. B
\textbf{37}, 785 (1988).

\bibitem{grimme_d3}
S. Grimme, J. Antony, S. Ehrlich and H. Krieg,
J. Chem. Phys.
\textbf{132}, 154104 (2010).

\bibitem{janz88}
G. J. Janz.
{\it Thermodynamic and Transport Properties for Molten Salts:  Correlation Equations for Critically Evaluated Density, Surface Tension,
Electrical Conductance and Viscosity Data.}
J. Phys. Chem. Ref. Data
\textbf{17}, Suppl. 2 (1988).

\bibitem{bussi2007}
G. Bussi, D. Donadio and M. Parrinello,
J. Chem. Phys.
\textbf{126}, 014101 (2007).

\bibitem{tissen94}
J. T. W. M. Tissen, G. J. M. Janssen and P. van der Eerden,
Mol. Phys.
\textbf{82}, 101 (1994).

\bibitem{koishi00}
T. Koishi, S. Kawase, S. Tamaki and T. Ebisuzaki
J. Phys. Soc. Jpn.
\textbf{69}, 3291 (2000).

\bibitem{hoover85}
W. G. Hoover,
Phys. Rev. A
\textbf{31}, 1695 (1985).

\bibitem{melchionna93}
S. Melchionna, G. Ciccotti and B. L. Holian,
Mol. Phys.
\textbf{78}, 533 (1993).

\bibitem{dlpoly}
I. T. Todorov, W. Smith, K. Trachenko and M. T. Dove,
J. Mater. Chem.
\textbf{16}, 1911 (2006).

\bibitem{moldy}
K. Refson,
Comput. Phys. Commun.
\textbf{126}, 310 (2000).

\bibitem{birch47} F. Birch,
Phys. Rev.
\textbf{71}, 809 (1947).

\bibitem{janz82} G. J. Janz and N. P. Bansal,
J. Phys. Chem. Ref. Data
\textbf{11}, 505 (1982).

\bibitem{kanai2013}
Y. Kanai, K. Fukunaga, K. Terasaka and S. Fujioka,
Chem. Eng. Sci.
\textbf{100}, 153 (2013).

\bibitem{janz79} G. J. Janz {\it et al.}
{\it Physical Properties Data Compilations Relevant to Energy Storage}.
Nat. Stand. Ref. Data Ser., Nat. Bur. Stand. (U.S.) 61, part II (1979).

\bibitem{lair2012}
V. Lair, V. Albin, A. Ringued\'e and M. Cassir,
Int. J. Hydrogen Energy
\textbf{37}, 19357 (2012).

\bibitem{murch82}
G. E. Murch,
Solid State Ion.
\textbf{7}, 177 (1982).

\bibitem{kashyap11}
H. K. Kashyap, H. V. R. Annapureddy, O. Ranieri and C. J. Margulis,
J. Phys. Chem. B
\textbf{115}, 13212 (2011).

\end{thebibliography}

\end{document}